%% file: vldb21.tex
\documentclass[sigconf, nonacm]{acmart}
\usepackage[ruled,linesnumbered]{algorithm2e}
\usepackage{color}
\usepackage{listings}
\usepackage{subfigure}
\usepackage{multirow}
\usepackage{balance}

\SetKwFor{ParaFor}{parallel for}{do}{end}

\begin{document}
\title{Large-Scale Approximate \textit{k}-NN Graph Construction on GPU}

\author{Hui Wang}
\affiliation{%
  \institution{Xiamen University}
  \city{Xiamen}
  \country{China}
}
\email{hwang2019@stu.xmu.edu.cn}

\author{Wan-Lei Zhao}
\affiliation{%
  \institution{Xiamen University}
  \city{Xiamen}
  \country{China}
}
\email{wlzhao@xmu.edu.cn}

\author{Xiangxiang Zeng}
\affiliation{%
  \institution{Hunan University}
  \city{Changsha}
  \country{China}
}
\email{xzeng@hnu.edu.cn}

\begin{abstract}
\textit{k}-nearest neighbor graph is a key data structure in many disciplines such as manifold learning, machine learning and information retrieval, etc. NN-Descent was proposed as an effective solution for the graph construction problem. However, it cannot be directly transplanted to GPU due to the intensive memory accesses required in the approach. In this paper, NN-Descent has been redesigned to adapt to the GPU architecture. In particular, the number of memory accesses has been reduced significantly. The redesign fully exploits the parallelism of the GPU hardware. In the meantime, the genericness as well as the simplicity of NN-Descent are well-preserved. In addition, a simple but effective \textit{k}-NN graph merge approach is presented. It allows two graphs to be merged efficiently on GPUs. More importantly, it makes the construction of high-quality \textit{k}-NN graphs for out-of-GPU-memory datasets tractable. The results show that our approach is 100-250$\times$ faster than single-thread NN-Descent and is 2.5-5$\times$ faster than existing GPU-based approaches.
\end{abstract}

\maketitle


\input{intro.tex}
\input{relat.tex}
\input{pre.tex}
\input{gnnd.tex}
\input{smerge.tex}
\input{exp.tex}
\input{conc.tex}

\begin{acks}
This work is supported by National Natural Science Foundation of China under grants 61572408 and 61972326, and the grants of Xiamen University 20720180074.
\end{acks}

\bibliographystyle{ACM-Reference-Format}
\balance
\bibliography{vldb21}

\end{document}

%% file: intro.tex
\section{Introduction}
Given a dataset $S$, the \textit{k}-NN graph is a directed graph structure, in which each node is directed to its top-\textit{k} nearest neighbors in $S$ under a given distance metric. It is a key data structure in manifold learning~\cite{tenenbaum2000global, roweis2000nonlinear, belkin2001laplacian}, machine learning~\cite{boiman2008defense} and information retrieval~\cite{weidong}, etc. The time complexity of building a \textit{k}-NN graph is $O(d{\cdot}n^2)$ when it is undertaken in an exhaustive manner, which is prohibitively expensive for large-scale datasets. Due to the fundamental role of \textit{k}-NN graph in various areas, continuous efforts have been made on the exploration of efficient solutions. Since the time complexity is too high to build an exact \textit{k}-NN graph, most of the works in the literature~\cite{chen2009fast, weidong} focus on the construction of approximate \textit{k}-NN graph.

Several efficient approaches have been proposed in recent years. The classic approach NN-Descent was proposed in 2011. It turns out to be a simple but effective solution. According to the paper, only a small portion of the comparisons are required to build a high-quality \textit{k}-NN graph. It is a generic approach in the sense that it is feasible for various distance metrics. Recent approach EFANNA~\cite{fu2016efanna} adopts a similar graph refinement strategy as NN-Descent on the rough graphs that are built by the space partitioning strategy. EFANNA demonstrates considerably higher efficiency over NN-Descent. Typically, only a few minutes are required to construct a million-level graph by a single thread. Unfortunately, this scheme is only feasible for $\textit{l}_p$-norms.

Although both NN-Descent and EFANNA are efficient and easily parallelizable, they are designed for CPU only. As GPU-based computation gets increasingly popular in various fields, more and more researchers explore efficient solutions on GPUs. Thanks to its high performance in floating-point computation, many GPU-based approaches~\cite{garcia2008fast, garcia2010k, johnson2019billion, pan2011fast, wieschollek2016efficient, zhao2020song, groh2019ggnn} have been proposed for approximate nearest neighbor (ANN) search. Recent experiments show that building \textit{k}-NN graph exhaustively on the GPU~\cite{garcia2008fast, garcia2010k,johnson2019billion} is even much faster than that of NN-Descent on the CPU~\cite{weidong}. Whereas, it is impractical to transplant classic NN-Descent directly to GPU due to the apparent difference in computation architecture. It becomes necessary to explore the efficient solution for \textit{k}-NN graph construction on GPUs. Moreover, another critical issue has been long overlooked in the literature. Up-to-now, there is no solution about how to build a \textit{k}-NN graph when the data is too big to be loaded in the memory. This issue becomes more and more imminent in the era of big data.

In this paper, we explore how to construct the large-scale \textit{k}-NN graph on GPU. A GPU-based NN-Descent (\textit{GNND}) is presented. In general, GNND follows the major steps of NN-Descent. However, considerable modifications have been made to adapt it to GPU architecture. These modifications include the sampling strategy, distance calculation during cross-matching, and the strategies in graph update. All of these schemes in combination lead to considerably faster speed over the CPU version. Meanwhile, the genericness of the original NN-Descent is well-preserved in GNND. In addition, a simple but effective \textit{k}-NN graph merge approach, that is built upon GNND, is presented. It allows two graphs to be merged efficiently on GPU. More importantly, it makes the \textit{k}-NN graph construction for out-of-GPU-memory dataset tractable via a divide-and-conquer manner. 

The remainder of this paper is organized as follows. The related works are reviewed in Section~\ref{sec:relate}. In Section~\ref{sec:prel}, NN-Descent and the architecture of GPU are reviewed as the preliminaries of our presentation in the later sections. Section~\ref{sec:construction} presents our designs and optimizations of GNND running on a single GPU. In Section~\ref{sec:large-scale}, our solution for large-scale \textit{k}-NN graph construction is presented. The experimental studies about the effectiveness of the proposed approach are presented in Section~\ref{sec:experiments}. Section~\ref{sec:conclusion} concludes the paper.

%% file: relat.tex
\section{Related Work}
\label{sec:relate}
There are in general two categories of approaches for \textit{k}-NN graph construction. Approaches such as~\cite{chen2009fast, wang2012scalable, zhang2013fast} follow the divide-and-conquer strategy. The dataset is divided into multiple subsets. The samples in each subset are expected to be relatively close to each other. The division scheme has to be specifically designed. The existing partition approaches include the hierarchical random projections~\cite{wang2012scalable}, \textit{Recursive Lanczos Bisection}~\cite{chen2009fast} or a series of locality sensitive hash functions~\cite{zhang2013fast}. A sub-graph is constructed for each subset in a brute-force way. Finally, these sub-graphs are merged into one \textit{k}-NN graph. Most of these divide-and-conquer approaches are designed for $\textit{l}_p$-space as it is difficult to define a division scheme for other metric spaces.

NN-Descent~\cite{weidong} is the representative approach of the second category. The \textit{k}-NN graph is constructed by iterative refinement on a random \textit{k}-NN graph. For an object $u$ in the dataset, each iteration of NN-Descent introduces all the neighbors of $u$ to each other. Each neighbor $v$ of $u$ will try to find the neighbor of itself in the neighborhood of $u$. The discovered new neighbors of $v$ will be used to update the neighborhood of $v$. Since most of the operations of NN-Descent are performed in a local of an object, it has a good parallel performance on CPUs. In NN-Descent, the newly discovered neighbor will be used to update the neighborhood of an object immediately. Since many new neighbors are produced during the iteration, a lot of random memory accesses are inevitable. A more detailed review of NN-Descent is presented in the next section. Recently, the hybrid scheme based on both the space partitioning approach and NN-Descent has been proposed in the literature [6]. Although efficient, this hybrid scheme is infeasible for the metrics beyond $\textit{l}_p$-norms.

Due to the high performance of GPU, several \textit{k}-NN graph construction approaches based on GPUs are also seen in the literature. FAISS-GPU~\cite{johnson2019billion} uses a new design for \textit{k}-selection, making brute-force \textit{k}-NN search on GPUs much faster than previous implementations. In the same work, the product quantization (PQ)~\cite{jegou2010product} along with \textit{k}-selection is proposed to construct the \textit{k}-NN graph construction for the large-scale dataset. Whereas extra time costs are introduced for codebook training and quantization. Moreover, high-quality \textit{k}-NN graph cannot be expected due to the quantization loss. GGNN~\cite{groh2019ggnn} is another representative approach that works on GPU. It constructs \textit{k}-NN graph in hierarchy. GGNN first divides the dataset into several small subsets. The \textit{k}-NN graph for each subset is constructed exhaustively on GPU. A new upper layer is formed by sample objects from each subset on the bottom layer. The sample objects are divided into several subsets, on which a set of sub-graphs are constructed. A hierarchy is built as the sampling and construction continue until there is only one sub-graph on the layer. With the neighborhood relations from the upper layer, a united \textit{k}-NN graph can be built on the lower layer. As this operation propagates to the bottom layer, a complete \textit{k}-NN graph is built for the whole dataset. GGNN uses greedy best first search with backtracking to find the nearest neighbors, a large number of random accesses to the GPU memory becomes inevitable, which impacts its efficiency on GPU considerably.

%% file: pre.tex
\begin{figure*}[t]
    \centering
    \subfigure[Simplified hardware architecture of NVIDIA post-Volta GPU]
    {\includegraphics[width=0.405\linewidth]{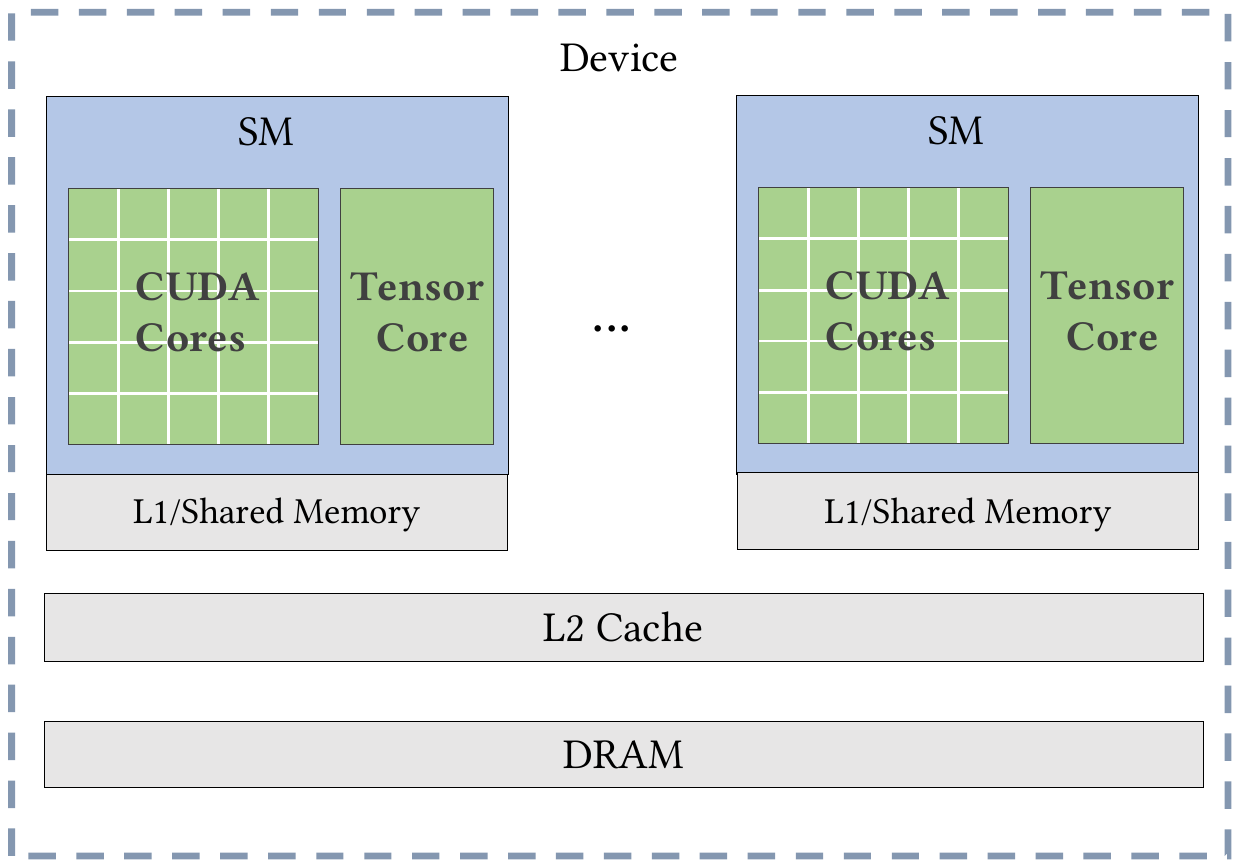}}
    \hspace{0.2in}
    \subfigure[Memory hierarchy of CUDA]
    {\includegraphics[width=0.39\linewidth]{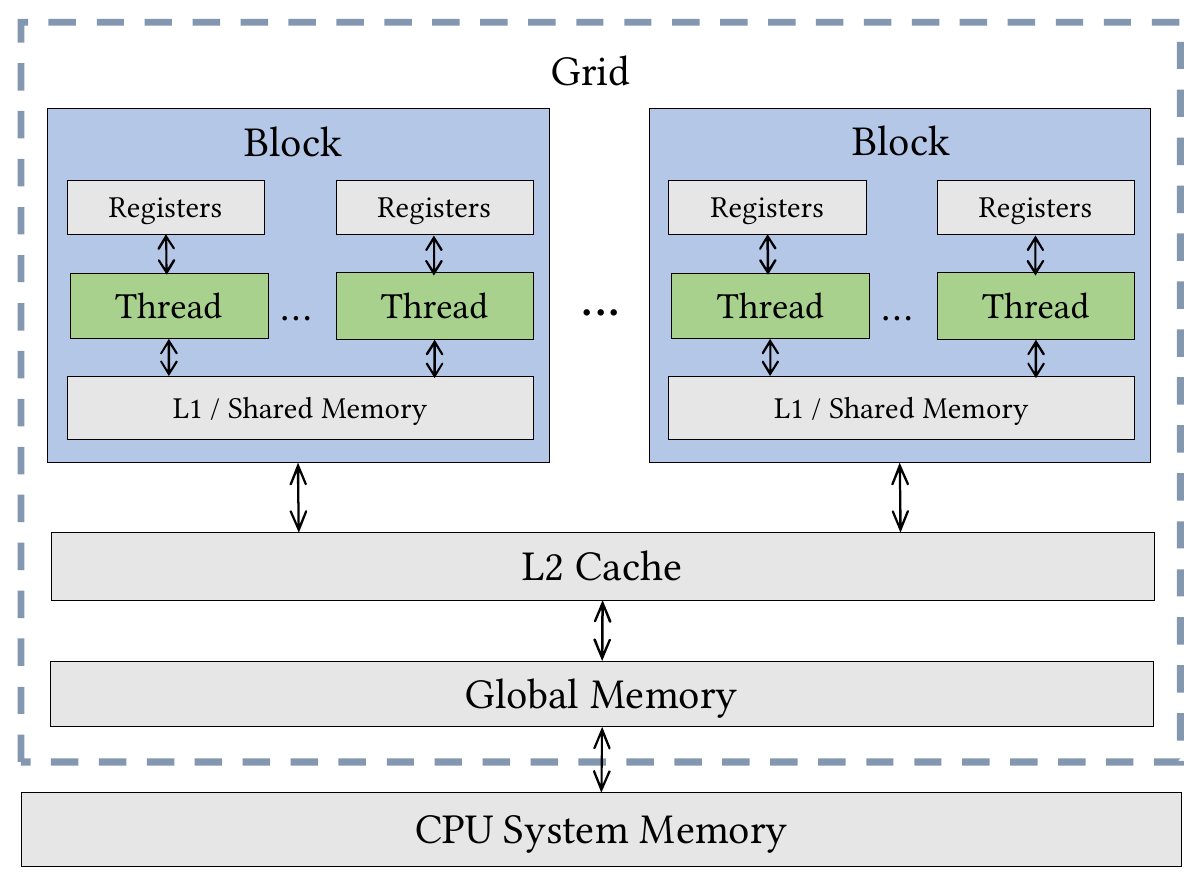}}
    \caption{Hardware architecture (figure a) and memory hierarchy (figure b) of an NVIDIA post-Volta GPU.}
    \label{fig:gpu}
\end{figure*}

\section{Preliminaries}
\label{sec:prel}

\subsection{NN-Descent on CPU}
NN-Descent~\cite{weidong} is known as a classic \textit{k}-NN graph construction approach. Until now, it is still the most efficient approach that works in generic metric space. The algorithm is built upon the principle of ``a neighbor of a neighbor is likely to be a neighbor''. In this section, a brief review of it is presented to support the discussions in the later sections. 

In general, four steps are involved in NN-Descent. 1. Initialize a \textit{k}-NN graph with \textit{k} random neighbors for each object; 2. Sample out a few \textit{new} objects and \textit{old} objects from within \textit{k}-NN list and reverse \textit{k}-NN list of one object. 3. Perform cross-matching on new-new object pairs as well as the new-old pairs of each object; 4. Insert the closer neighbors produced from \textit{Step-3} into the \textit{k}-NN list of corresponding objects. The last three steps are repeated for several rounds until it converges. At the first iteration, all the objects in one \textit{k}-NN list are viewed as \textit{new}. The objects that are newly inserted into a \textit{k}-NN list at \textit{Step-4} are labeled as \textit{new}. Otherwise, the objects in one \textit{k}-NN list are labeled as \textit{old}. According to~\cite{weidong}, it is no need to consider all the \textit{new} and \textit{old} objects in the \textit{k}-NN list and reverse \textit{k}-NN list of one object. As a result, sampling in \textit{Step-2} is required. The cross-matching in \textit{Step-3} computes the distances between \textit{new} objects and the distances between \textit{new} and \textit{old} objects. As a result, closer neighbors are produced for the objects joined in the comparison, which are used to update the corresponding NN lists. Since far neighbors are replaced by the closer neighbors after the insertion each time, the \textit{k}-NN graph evolves to better quality monotonically.

Objects in the \textit{k}-NN lists are driven by the cross-matching to reach to the closer neighbors in each iteration. Essentially, NN-Descent is a hill-climbing algorithm undertaken in batch~\cite{weidong}. This approach works particularly well when the data dimension or the intrinsic data dimension is low. When NN-Descent runs on CPU, more than \textit{90\%} of the computation has been spent on the distance calculation in \textit{Step-3}.

\subsection{GPU Structures}
In this section, we briefly introduce the architecture and programming model of NVIDIA GPU for general-purpose parallel computing. An NVIDIA GPU consists of dozens of streaming multiprocessors (SMs). Each SM contains dozens of processing cores and can execute hundreds of threads concurrently. The simplified hardware architecture of NVIDIA post-Volta GPU is shown in \autoref{fig:gpu}(a).

A function that is to be run on the GPU is called as a kernel. A grid of thread blocks are launched for one kernel. All the threads of one kernel execute the same code. One block is physically corresponding to one SM. A block in an SM is further divided into sub-blocks called warps as the unit of thread scheduling. In CUDA, one warp is comprised of \textit{32} threads. All the threads in a warp follow the Single Instruction, Multiple Thread (SIMT) model. With the support of independent thread scheduling of NVIDIA's post-Volta GPU, diverged threads could run on the same warp, which is much more flexible.

The memory hierarchy is shown in \autoref{fig:gpu}(b). Registers are private to each thread. Threads in the same warp could exchange variables in the registers by using warp shuffle functions. Inside each SM, there is a fast on-chip memory. This memory can be used as either L1 cache or shared memory. The latter is visible to the programmers. The threads in one block are able to communicate with each other via this shared memory. Both registers and shared memory are very efficient to access however are of limited capacity. In contrast, global memory (physically DRAM) has much larger capacity, whereas has higher latency, which becomes the bottleneck for memory access intensive tasks.

The advantage of GPU-based computation over CPU lies in its high parallelization. This is the main reason that the exhaustive \textit{k}-NN graph  construction on GPU is considerably faster than building \textit{k}-NN graph by NN-Descent on CPU. This motivates us to run NN-Descent on GPU. However, due to the intensive memory access in this algorithm both for distance computation and data insertion, it is non-trivial to transplant the NN-Descent from CPU to GPU. There are several aspects to be considered. First of all, in order to make full use of the GPU's floating-point computing capability, the way of memory access should be optimized. It would be more memory-friendly if all the threads in a warp load contiguous data from the memory. Due to the latency mismatch between GPU core and global memory, it is better to reduce the frequency of accessing to the global memory. 

In our design, the computation power of GPU as well as the latency mismatch between different hierarchies of GPU memory are carefully considered. Considerable modifications have been made to the original NN-Descent algorithm. Specifically, the modifications include the sampling strategy on the objects to be joined in the next round cross-matching, the distance computation between object pairs, and the insertion strategy for the produced closer neighbors. All of these strategies in combination lead to significantly faster construction speed while maintaining high-quality of the \textit{k}-NN graph.

%% file: gnnd.tex
\section{Approximate \textit{k}-NN Graph Construction on GPU}
\label{sec:construction}
In this section, an efficient \textit{k}-NN graph construction approach on GPU is presented. Basically, we follow the major steps in the classic NN-Descent. The presentation of this section focuses on the major modifications over NN-Descent.

In NN-Descent, distance calculations take up most of the construction time. In contrast, the distance calculation is no longer the processing bottleneck on GPU due to its superior floating-point performance. Because of the latency mismatch between different hierarchies of the memory, the frequent update operations in NN-Descent, which requires intensive memory accesses, become the processing bottleneck. Moreover, the parallel update operations also induce frequent locks and unlocks on each \textit{k}-NN list. As a result, an additional delay is introduced to the whole construction process. The design of our construction algorithm considers how to make full advantage of high GPU floating-point performance and alleviate the delay between the GPU core and memory access in the meantime.

Given a dataset $S$ with $n$ objects, the \textit{k}-NN graph of dataset $S$ is a directed graph with a fixed degree of $k$. For each object $u \in S$, the \textit{k}-NN list of $u$ keeps its top-\textit{k} nearest neighbors. These \textit{k}-NN lists are sequentially stored in the global memory. Before the iteration starts, all the neighbors in the \textit{k}-NN list of $u$ are marked as \textit{NEW}. After the cross-matching and the update on each \textit{k}-NN list, the neighbors in one \textit{k}-NN list can be divided into two groups. The neighbors that are already in the list in the last iteration are called old neighbors. They are marked as \textit{OLD}. The neighbors that are newly inserted in the current iteration are called new neighbors. They are marked as \textit{NEW}. In the following, $u$'s neighbors marked with \textit{NEW} and \textit{OLD} are called $u$'s \textit{NEW} neighbor and \textit{OLD} neighbor, respectively. The sampled \textit{NEW} and \textit{OLD} neighbors are called \textit{NEW} sample and \textit{OLD} sample, respectively. 

\begin{algorithm}[htb]
  \caption{ConstructKNNGraph}
  \label{alg:gnnd}
  \KwIn{reference set $S$, number of NN \textit{k}, sample number $p$,  the maximum number of iterations $MaxIter$;}
  \KwOut{\textit{k}-NN graph $G$}

  \ParaFor{$s \in S$}
  {  
    $G[s]$ $\leftarrow$ $p$ random objects from \textit{S} \;
    Sort the objects in $G[s]$ in ascending order \;
    Mark the objects in $G[s]$ as \textit{NEW} \;
  }

  $t$ $\leftarrow$ MaxIter\;
  \While{$t > 0$} 
  {
    $G_{old}$, $G_{new}$ $\leftarrow$ ParallelSample($S$, $G$, $p$) \;
    \label{SampleInLine}
    \ParaFor{$s \in S$}
    {
      $S_{new}$ $\leftarrow$ $NEW$ samples in $G_{new}$[s] \;
      $D$ $\leftarrow$ CalculateDistances($S_{new}$) \;
      \ParaFor{$u \in S_{new}$}
      {
        $D_u$ $\leftarrow$ distances from other $NEW$ samples to $u$ in $D$\;
        $V$ $\leftarrow$ other $NEW$ samples \;
        ($v, d$) $\leftarrow$ GetNearestObject($u, V, D_u$) \;
        InsertIntoNNList($G[u], v, d$) \;
      }

      $S_{old}$ $\leftarrow$ $OLD$ samples in $G_{old}$[s] \;
      $D$ $\leftarrow$ CalculateDistances($S_{new}$, $S_{old}$) \;
      \ParaFor{$u \in S_{new}$}
      {
        $D_u$ $\leftarrow$ distances from $OLD$ samples to $u$ in $D$ \;
        $V$ $\leftarrow$ $OLD$ samples \;
        ($v, d$) $\leftarrow$ GetNearestObject($u, V, D_u$) \;
        InsertIntoNNList($G[u], v, d$) \;
      }
      \ParaFor{$u \in S_{old}$}
      {
        $D_u$ $\leftarrow$ distances from $NEW$ samples to $u$ in $D$\;
        $V$ $\leftarrow$ $NEW$ samples \;
        ($v, d$) $\leftarrow$ GetNearestObject($u, V, D_u$) \;
        InsertIntoNNList($G[u], v, d$) \;
      }
      Mark all sampled neighbors as $OLD$ \;
    }
    $t \leftarrow t - 1$\;
  }
  \Return{\textit{k}-NN graph $G$} \;
\end{algorithm} 

\subsection{Sampling on Close Neighbors}
\label{sec:smpl}
In NN-Descent iteration, a \textit{k}-NN list typically consists of OLD neighbors and NEW neighbors after one round of cross-matching and the  afterward graph update. The number of \textit{OLD} neighbors and \textit{NEW} neighbors varies from one list to another. Moreover, the numbers of \textit{OLD} neighbors and \textit{NEW} neighbors in the reverse \textit{k}-NN list are unpredictable, which could vary from a few to a few hundreds. If all of these \textit{OLD} and \textit{NEW} neighbors are selected to join the next round of cross-matching, these samples should be collected into a dynamic array for each \textit{k}-NN list. However, the cost of maintaining \textit{n} dynamic arrays is prohibitively high given the fact that the dynamic memory allocation is of high latency and the global memory might not be sufficient. To address this issue, only the first \textit{p} ($p < k$) objects are sampled for either \textit{OLD} neighbors or \textit{NEW} neighbors from one \textit{k}-NN list. This results in two adjacency graphs $G_{new}$ and $G_{old}$ in which the sampled $p$ \textit{NEW} neighbors and $p$ \textit{OLD} neighbors are kept for each object separately.

Given graph $G_{new}$ is constructed, each list in $G_{new}$ is visited to append the reverse neighbors derived from $G_{new}$ itself. Namely, given sample $v$ in $G_{new}[s]$, the list of $G_{new}[v]$ is appended with $s$. The appending is an atomic operation. It will be no longer undertaken as long as the size of $G_{new}[v]$ list reaches the upper bound $2p$. The same operation is carried out on the adjacency graph $G_{old}$. As the procedure is running as multi-thread on GPU,  this part of the operation can be regarded as random sampling. Since the number of sampled neighbors is fixed, we can store the results of the sampling in the form of fixed-degree graph (fixed-degree adjacency list). An extra array sized of \textit{n} is defined to keep the sizes of $G_{new}$ and  $G_{old}$ lists since not all the lists' sizes could reach $2p$.

In order to reduce the unnecessary distance calculations, we remove duplicates for each list from the resulting graphs $G_{new}$ and $G_{old}$. We simply assign a warp (32 threads) to each list for sorting, and then remove duplicate neighbors from each list. Although this is not the most efficient way on GPU, this simple way is preferred as the time cost of this operation is negligible.

It is also worth noting that since both $k$ and the number of samples $p$ are small, all the neighbors in a \textit{k}-NN list can be loaded into the shared memory for sampling. This will reduce the accesses to the global memory. In addition, reading and writing to global memory will be coalesced, which is very efficient on GPU as well. 


\subsection{Cross-matching on Sampled Neighbors}
\label{sec:crss}

\begin{figure}[t]
  \centering
  \includegraphics[width=\linewidth]{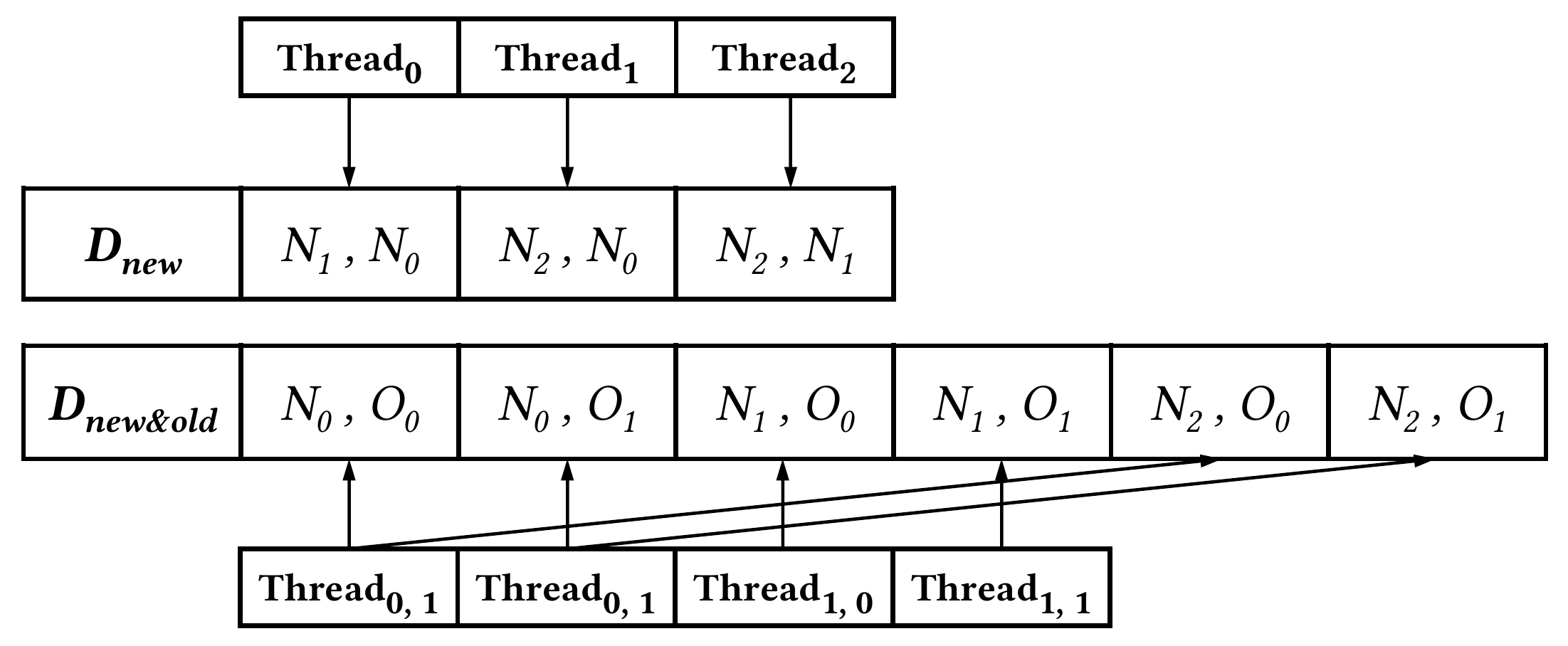}
  \caption{An example to show the arrangement of distance calculation results in the shared memory and the corresponding thread for calculating each result. $N_i$ represents the \textit{i}-th \textit{NEW} samples in the \textit{k}-NN list, $O_i$ represents the \textit{i}-th \textit{OLD} samples in the \textit{k}-NN list.}
  \label{fig:dist}
\end{figure}

\begin{figure*}[t]
  \begin{center}
    \subfigure[Distance calculation on tiled]
    {\includegraphics[width=0.25\linewidth]{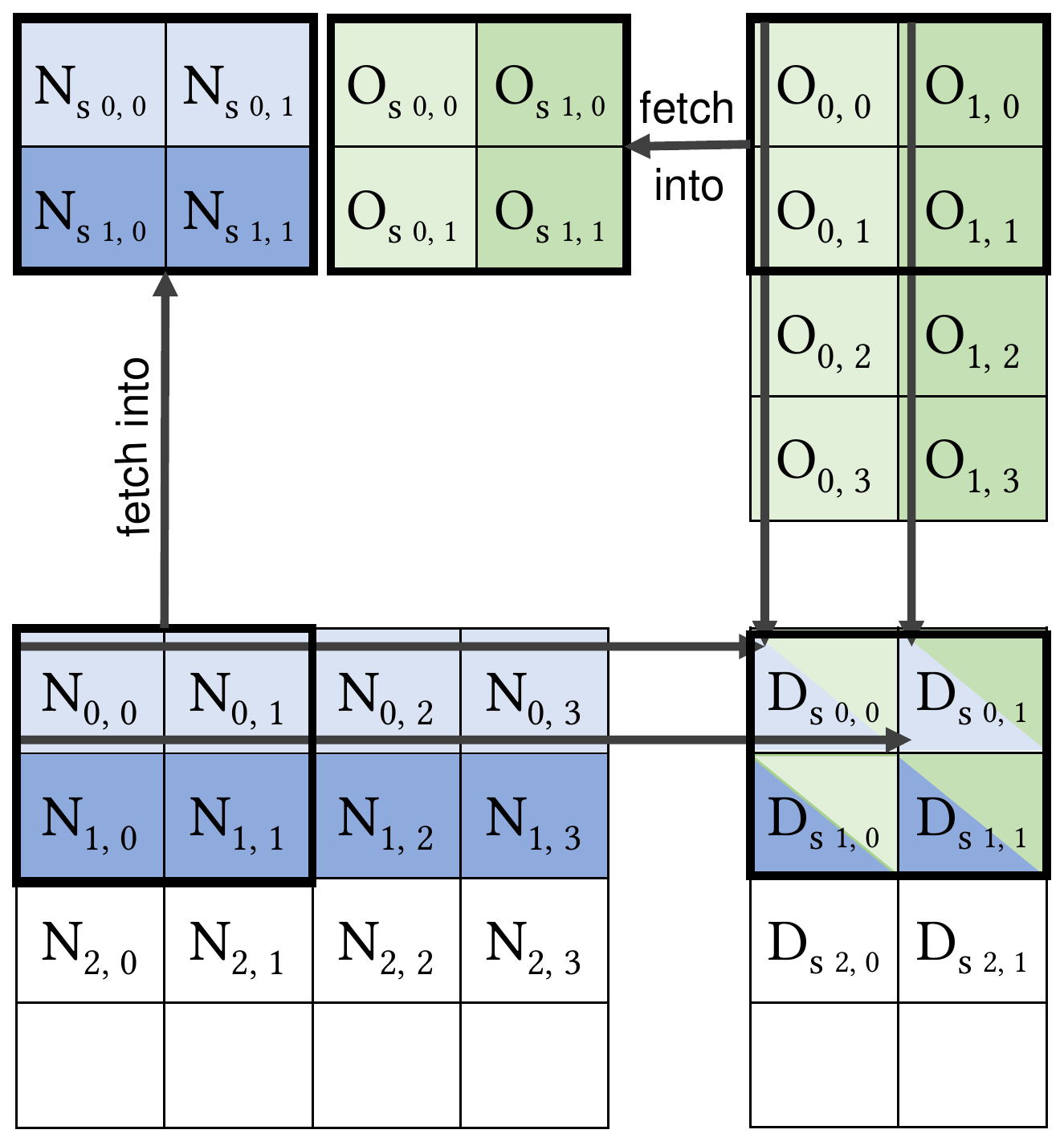}}
    \hspace{0.7in}
    \subfigure[Two phases distance calculation on tiled]
    {\includegraphics[width=0.42\linewidth]{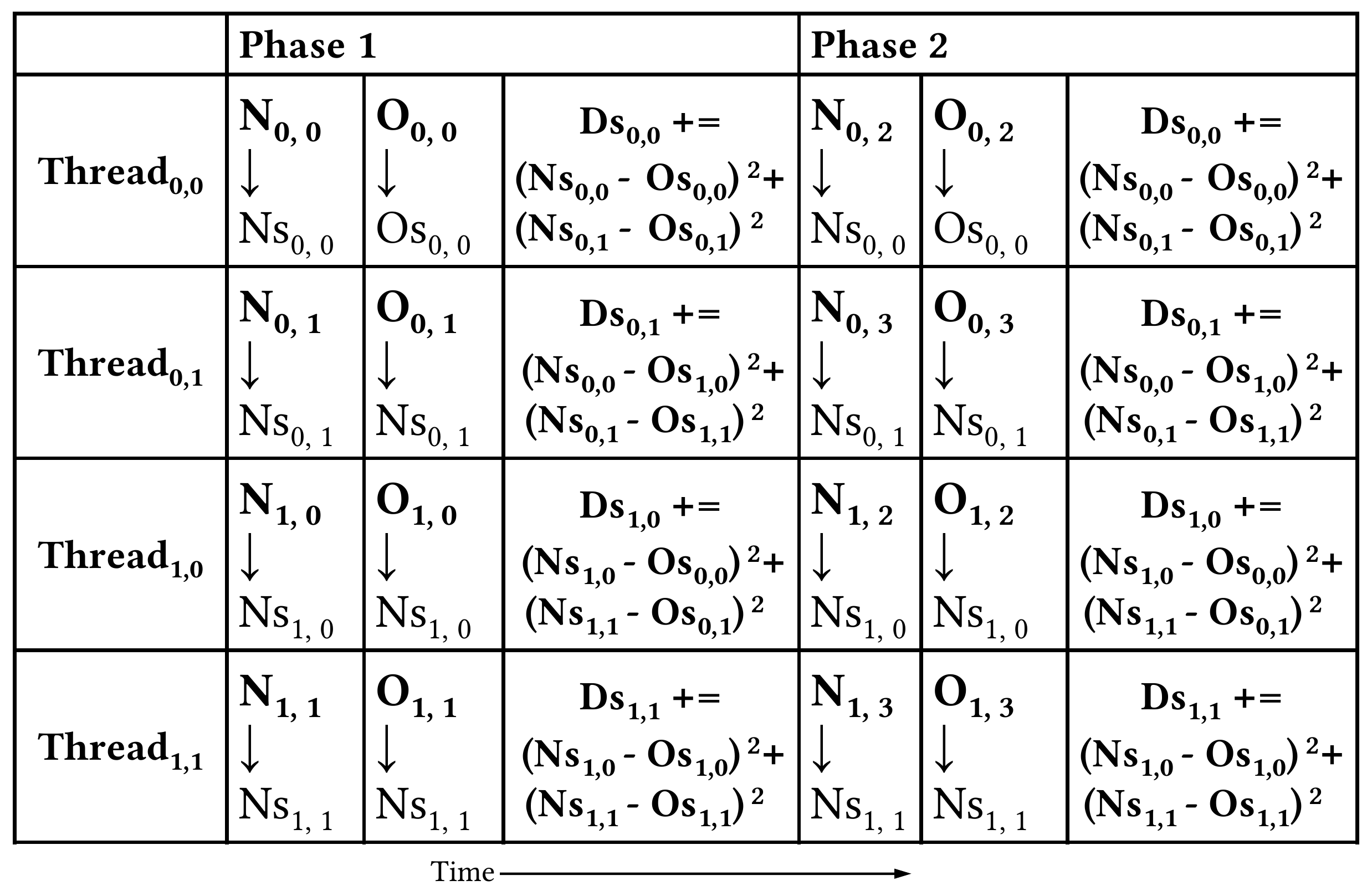}}
    \caption{The illustration of execution phases of tiled distance calculation on a $2 \times 2$ thread block. The subscript \textit{s} of the variables in the figures indicates that they are defined in the shared memory. In figure (a), the value of the $j$-th dimension of the $i$-th \textit{NEW} sample as $N_{i,j}$. Similarly, $O_{i,j}$ is the value of the $j$-th dimension of the $i$-th \textit{OLD} sample. In figure (b), the ``$\downarrow$'' represents that loading one dimension data from global memory to the shared memory. Data in shared memory are accessible for all the threads in one block. Figure (a) illustrates the Phase 1 of figure (b). Figure (b) illustrates the two phases that calculates the distances between two \textit{NEW} samples and two \textit{OLD} samples. The tiles (thick line boxes) in figure (a) will slide to calculate all distances between three \textit{NEW} and two \textit{OLD} samples.}
    \label{fig:matmult_proc}
  \end{center}
\end{figure*}

As the \textit{NEW} and \textit{OLD} samples are ready, the next step is to perform cross-matching on the sampled lists. Namely,  our task is to calculate the distances between the \textit{NEW} samples and the distances between the \textit{NEW} samples and the \textit{OLD} samples. The results of the calculations are kept in the shared memory. Two continuous blocks in the shared memory are allocated to keep these distances for \textit{NEW}-\textit{NEW} pairs and the \textit{NEW}-\textit{OLD} pairs respectively. As the number of \textit{NEW} samples and the number of \textit{OLD} samples are no bigger than $2p$, the size of these two blocks is small and in fixed length.

For the \textit{NEW}-\textit{NEW} pairs and \textit{NEW}-\textit{OLD} pairs, two different distance calculation strategies are used in our approach. For the \textit{NEW} samples of an object, we use a thread block to perform the distance calculations. Since the number of the \textit{NEW} samples is usually small, all the vectors are loaded into the shared memory. \autoref{fig:dist} shows their arrangement in the shared memory. If the data dimension is too high that cannot be loaded into the shared memory at once, they are divided into multiple sub-vectors. The distance calculation is carried out on the sub-vectors. The resulting distances are further reduced into the overall distance between two vectors. Given samples $u$ and $v$, the distance between them is kept at position $ \frac{u(u-1)}{2} + v$ of $D_{new}$ in the shared memory. We assign each thread in the thread block with the distance calculation task between two samples. The samples $u$ and $v$ to be calculated by thread-$t$ are obtained as 
\begin{equation}
  u = \lceil \sqrt{2t + 2.25} - 0.5 \rceil,
\end{equation}
\begin{equation}
  v = t - \frac{u(u-1)}{2}.
\end{equation}
Due to the limited number of threads in the thread block (\textit{1024} in CUDA), some threads are called for several rounds. For instance, given there are \textit{512} threads in one block, thread-0 should not only calculate $D_{new}$[0], but also $D_{new}$[512] and $D_{new}$[1024]. This can be achieved by adding an offset to $t$ in the above equations.

For the distance calculation between the \textit{NEW} and the \textit{OLD} samples, it is treated as a matrix multiplication problem. A variant of tiled matrix multiplication is adopted to calculate the distances between the \textit{NEW} samples and the \textit{OLD} samples. The major difference from the standard tiled matrix multiplication~\cite{kirk2016programming} is that the dot product operation in matrix multiplication is replaced with the specified distance metric, such as $\textit{l}_2$-norm and \textit{K}-Square.

\autoref{fig:matmult_proc} illustrates how the $\textit{l}_2$-norm is calculated between three \textit{NEW} samples and two \textit{OLD} samples (all of which are \textit{4}-d vectors). In the illustration, we assume there are only four threads ($2{\times}2$ tiled) in a GPU block.  Moreover, we assume there are only three $2{\times}2$ cells in the shared memory. One cell could keep a single floating-point value. These $2{\times}2$ cells are accessible by all the threads in this block. At the beginning of one calculation, the dimension values from a pair of \textit{OLD} and \textit{NEW} samples are loaded into these $2{\times}2$ cells from global memory. As the cells could not hold all the dimensional values (i.e., \textit{8} values) of two vectors at once, the distance calculation on a sample pair is divided into two phases in this example. At the first phase, the first two dimensions of two vectors are loaded, the $\textit{l}_2$-norm between the sub-vectors is calculated and aggregated. At the second phase, the $\textit{l}_2$-norm between the remaining two dimensions is calculated and aggregated. It should be noted that the vectors of these samples are stored separately in the global memory, we do not assemble them into a matrix in advance. For this reason, popular libraries such as cuBLAS are not employed. \autoref{fig:matmult_proc} details the sequence of reading the global memory values into the shared memory.
 
In practice, we use $16 \times 16$ tiles, and we only use one thread block to perform distance calculations for an object by traversing all the tiles. It is enough to make full use of GPU, because all the distance calculations for all the objects are performed at the same time.
 
\subsection{\textit{k}-NN Graph Update}
After the cross-matching on the \textit{NEW}-\textit{NEW} sample pairs and the \textit{NEW}-\textit{OLD} sample pairs, distances that are associated with the corresponding sample pairs are kept  in the  shared memory. In classic NN-Descent, all these produced neighbor pairs are used to update the \textit{k}-NN graph. However, this is too costly in our GPU implementation since the access to the global memory is much slower than the distance calculation on GPU cores. Additionally, locks on the \textit{k}-NN lists are inevitable when a large  number of produced pairs are inserted into the graph simultaneously. This slows down the whole process further. As a result, the graph update becomes the processing bottleneck. In this section, two schemes are proposed to address this issue. Firstly, only a few produced neighbors are used to update the graph. The time consumption due to graph update is further reduced by setting multiple spinlocks on different segments of a \textit{k}-NN list, which allows the insertion to be fulfilled in parallel on one list. 

\textbf{Selective update} Different from classic NN-Descent, only very few produced neighbor pairs for one object are selected to update the graph in our approach. In an object local, for a \textit{NEW} sample \textit{u}, we only use the nearest object of \textit{u} from the other \textit{NEW} samples and the nearest object of \textit{u} from the \textit{OLD} samples to update the \textit{k}-NN list of \textit{u}. As for an \textit{OLD} sample, we select the nearest object of it from \textit{NEW} samples to update its \textit{k}-NN list. Algorithm~\ref{alg:choose_insertion} summarizes the process to find the nearest object for \textit{u}. Every \textit{32} objects that have calculated distance from \textit{u} are handled by a warp. If the number of objects is less than \textit{32}, the remaining part will be filled with $(\infty, \infty)$. A standard shuffle based minimal reduction operation (\autoref{alg:choose_insertion} \textit{Lines}~\ref{reduction_begin}--\ref{reduction_end}) on the \textit{32} objects is performed to find the nearest sample of \textit{u}. Function \textit{\_\_shfl\_down(var, delta)} will copy a variable (\textit{var}) from a thread in the same warp, the second parameter \textit{delta} is the difference between the ID of the target thread and the ID of the caller thread. The warp-level parallel reduction can make full use of the hardware of GPU, which makes the above operations have excellent performance. After the reduction is executed, the nearest object selected by each warp of one group is stored in the variable of \textit{thread-0} of each warp. In the end, the resulting tuple to return will be updated by the variable of \textit{thread-0}.

\begin{algorithm}[htb]
  \caption{GetNearestObject($\cdot$)}
  \label{alg:choose_insertion}
  \KwIn
  {
    sample \textit{u}, sample set \textit{V} in which all the distances of samples have been calculated from \textit{u}, distance calculation results $D_{u}$;
  }
  \KwOut
  {
    a tuple $(v, d)$ which consists of the nearest object $v$ and the distance from $v$ to $u$.
  }
  $(v, d) \leftarrow (\infty, \infty)$ \;
  \ParaFor{every 32 objects $V_{sub} \subseteq V$}
  {
    \ParaFor{thread t in a warp}
    {
      \eIf{$V_{sub}[t]$ exists}
      {
        $v_{t} \leftarrow V_{sub}[t]$ \;
        $d_{t} \leftarrow D_{u}[v_{t}]$ \;
      }
      {
        $v_{t} \leftarrow \infty$ \;
        $d_{t} \leftarrow \infty$ \;
      }
      $i \leftarrow 16$ \;
      \label{reduction_begin}
      \While{$i > 0$}
      {
        $(v_{t}, d_{t})$ $\leftarrow$ min($(v_{t}, d_{t})$, \_\_shfl\_down($(v_{t}, d_{t}), i)$) ;
        /* $(v_{a}, d_{a}) < (v_{b}, d_{b})$ when $d_{a} < d_{b}$  */ \\
        $i \leftarrow i / 2$ \;
      }
      \label{reduction_end}
      \If{$t == 0$}
      {
        $(v, d)$ = atomicMin($(v, d)$, $(v_{t}, d_{t})$) \;
      }
    }
  }
  return $(v, d)$ \;
\end{algorithm} 

\textbf{Multiple Spinlocks} Typically, three produced neighbor pairs associated with distances are returned by \autoref{alg:choose_insertion} for one \textit{k}-NN list. They are ready to be inserted into corresponding \textit{k}-NN lists. Threads access to the \textit{k}-NN list in parallel to find the position where object \textit{v} is to be inserted. Since \textit{k} is usually small, the above operation can be performed directly in the registers. The neighbors in the \textit{k}-NN list are stored sequentially in the global memory, the accesses to the global memory will be coalesced. As only very few neighbors are selected to insert and the insertion is undertaken in parallel, the graph update is much faster than inserting all the produced neighbor pairs.

In order to maintain the data integrity, a spinlock should be imposed on the \textit{k}-NN list during an insertion. This is, however, inefficient on GPU. In order to alleviate the latency caused by the spinlocks, the \textit{k}-NN list is divided into multiple segments. Therefore, the spinlock is imposed on the sub-list level, which allows the parallel insertions on one \textit{k}-NN list. Additionally, locks on the sub-list level also reduce the number of threads and global memory accesses required for each insertion. For instance, \textit{k}-NN list is divided into $k / 32$ segments. Each segment keeps \textit{32} (the size of a warp) neighbors. The object \textit{v} will be inserted into the $v \% (k / 32)$-th segment. By applying the optimization, each spinlock only needs to guard one segment, which allows multiple insertion operations to be performed on a \textit{k}-NN list at the same time, thereby reduces the possible waiting time of each thread. In addition, since the number of samples in each segment is the same as the warp size, we only need one warp for one insertion. As a result, much fewer GPU resources are required. As the iteration is completed, all the segments of one \textit{k}-NN list will be merged into one. The speed-up brought by this \textit{multiple spinlocks} will be shown in the ablation study in the experiment section.

The complete GPU-based NN-Descent is presented in~\autoref{alg:gnnd}. Three major modifications over classic NN-Descent are presented. Firstly, only fixed number of \textit{NEW} samples and \textit{OLD} samples are selected for cross-matching (\textit{Line 8}). During cross-matching, the distances between \textit{NEW} samples and distances between \textit{OLD} samples and \textit{NEW} samples are calculated in different ways to make full use of GPU computation power (\textit{Lines 11} and \textit{19}). Finally, only the discovered close neighbors are selected to update the graph for speed efficiency (\textit{Lines 15}, \textit{23} and \textit{29}). 

\begin{figure}[t]
	\begin{center}
    \includegraphics[width=0.65\linewidth]{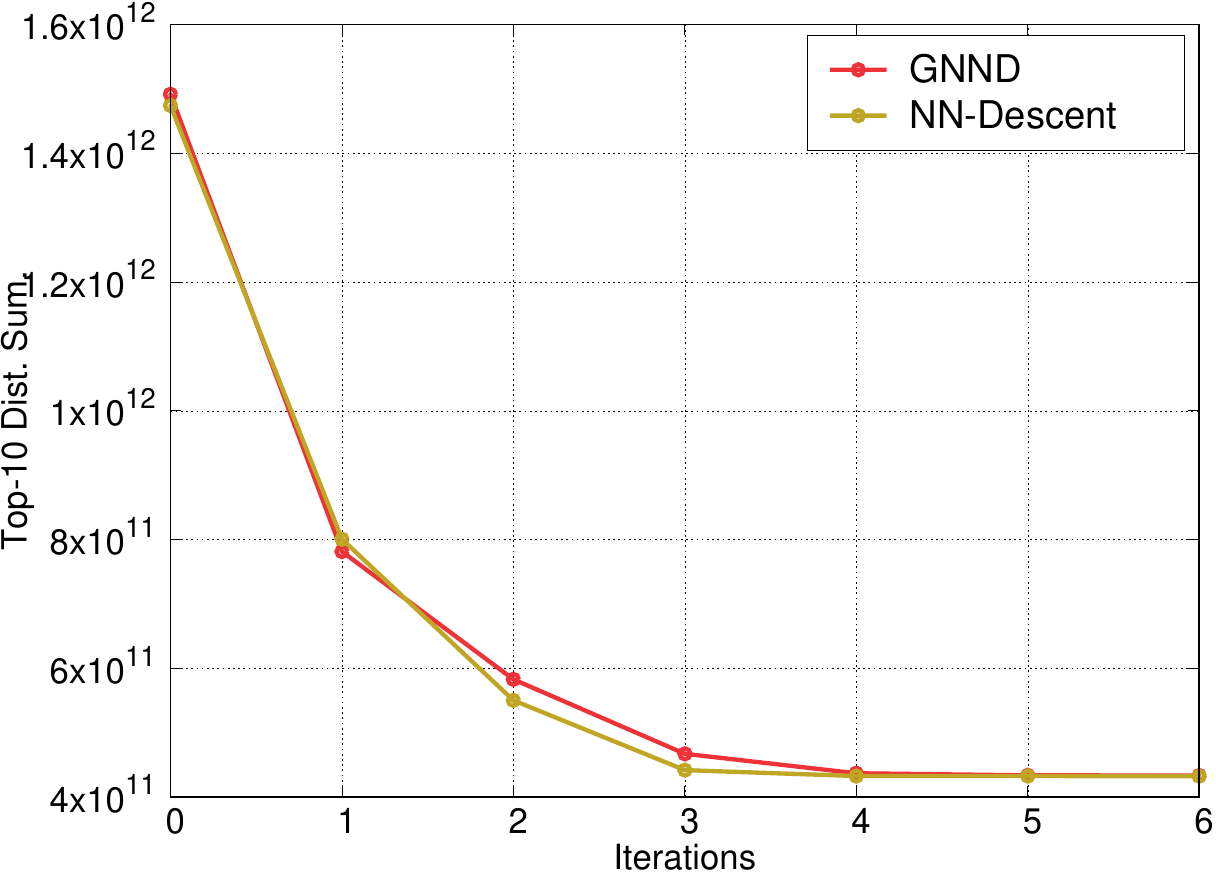}
    \caption{The variation trend of $\phi(G)$ of GNND in comparison to that of NN-Descent. \textit{k} is fixed to \textit{10} in the experiment.   \label{fig:dist_sum}}
\end{center}
\end{figure}

It is imaginable that the quality of \textit{k}-NN graph gets better after one round iteration since far neighbors are replaced by close neighbors on the \textit{k}-NN lists. However it is possible that the quality improvement might not be as fast as classic NN-Descent as many close neighbors are not selected to insert. To confirm the validity of our approach, we study the variation trend of the overall graph structure. A function $\phi(\cdot)$ is defined to measure the overall graph structure
\begin{equation}
\phi(G)=\sum_{i=1}^{n}\sum_{j=1}^{k}G_{ij},
\label{eqn:sum}
\end{equation}
where $G_{ij}$ returns the distance of the \textit{j}-th neighbor to object \textit{i}. The study is conducted on SIFT1M dataset. $\phi(G)$ is calculated after one round of iteration. The comparison is conducted between NN-Descent and GNND. As shown in~\autoref{fig:dist_sum}, the variation trend of GNND largely overlaps with that of NN-Descent. A similar trend is observed on other datasets, which implies it is sufficient to use only the closest neighbors to update the \textit{k}-NN list. The other relative far neighbors are quite likely squeezed out later. As a result, ignoring these neighbors during insertion has very little impact on the convergence speed.

%% file: smerge.tex
\section{GNND for Large-scale Datasets}
\label{sec:large-scale}
To this end, the proposed algorithm is able to build high-quality \textit{k}-NN graph efficiently. For instance, it only takes a few seconds for million level dataset, which is several hundreds times faster than the classic NN-Descent. However, the problem of \textit{k}-NN graph construction remains challenging given that we have billions of data to be processed. Neither the global memory in GPU nor the system memory could hold such large-scale data at once. Moreover, the large-scale data may not come at once, the \textit{k}-NN graph is required to be constructed incrementally in such scenario. In this section, a simple but effective \textit{k}-NN graph merge algorithm that builds on GNND is presented. Based on the merge algorithm, both the construction of very large-scale \textit{k}-NN graph and the incremental construction of \textit{k}-NN graph are easily tractable. Additionally, it allows the \textit{k}-NN graph construction to be carried out on multiple GPUs.

\subsection{\textit{k}-NN Graph Merge}
Given two \textit{k}-NN graphs $G_{1}$ and $G_{2}$ which are constructed from datasets $S_{1}$ and $S_{2}$ respectively\footnote{Without the loss of generality, we assume $S_{1} \bigcap S_{2} = \emptyset$.}, the task of \textit{k}-NN graph merge is to build graph $G$ for dataset $S = S_{1} \cup S_{2}$ based on $G_{1}$ and $G_{2}$. Intuitively, the graph merge on GPU can be fulfilled by NN search approaches proposed in SONG~\cite{zhao2020song} or GGNN~\cite{groh2019ggnn}. Namely, graph $G$ can be constructed by querying against graph $G_2$ with samples from $S_1$ and repeat the NN search alternatively between $S_2$ and $G_1$. Although it is feasible, the computation power of GPUs is not fully capitalized. 

In this paper, the \textit{k}-NN graph merge problem is addressed as a graph construction problem on a half-baked graph. Given datasets $S_{1}$ and $S_{2}$ with $n$ and $m$ objects respectively, we join their \textit{k}-NN graphs $G_1$ and $G_2$ directly into a graph $G$, which consists of $m + n$ \textit{k}-NN lists. For the first $n$ \textit{k}-NN lists in graph $G$, they actually belong to $G_1$. The last $\frac{k}{2}$ objects of one list from this sub-graph are replaced  with $\frac{k}{2}$  random objects from $S_{2}$. Similarly, for the last $m$ \textit{k}-NN lists in graph $G$, we replace the last $\frac{k}{2}$ objects of each list with $\frac{k}{2}$ random objects in $S_{1}$. After the initialization, GNND is called to refine graph $G$. Different from performing GNND on a raw set, we only need to perform cross-matching on the pairs from the different sub-graphs as the sub-graphs $G_1$ and $G_2$ are fully baked. Specifically, the last $\frac{k}{2}$ objects of each list are marked as \textit{NEW} samples initially. The distances between \textit{NEW} samples will not be calculated during GNND iteration. As a result, GNND turns out to be much faster than it runs on a raw set. This merge procedure is called GPU-based graph merge (GGM) and summarized in~\autoref{alg:smerge}.
 
\begin{algorithm}
  \KwData{dataset $S$, \textit{k}-NN Graph $G_{1}$ of $S_1$ of size \textit{m},  \textit{k}-NN Graph $G_{2}$ of $S_2$ of size \textit{n}}
  \KwResult{\textit{k}-NN Graph $G$}
  \textbf{Divide} $G_{1}$ into $G^{u}_{1}$ and $G^{v}_{1}$\;
  \textbf{Divide} $G_{2}$ into $G^{u}_{2}$ and $G^{v}_{2}$\;
  \For{i=1; i $\leq$ m; i++}
 {	 
    \textbf{Append} $G^{u}_{1}[i]$ with $\frac{k}{2}$ random samples from $S_2$\;
}
\For{i=1; i $\leq$ n; i++}
{
    \textbf{Append} $G^{u}_{2}[i]$ with $\frac{k}{2}$ random samples from $S_1$\;
}
$G \leftarrow G^{u}_{1} \cup G^{u}_{2} $\;

Call GNND to refine G\;
/* GNND is modified to only calculate the distances between objects in different subsets */ \\
$G$ $\leftarrow$ \textbf{Merge and Sort} $G$ with $G^{v}_{1}$ and $G^{v}_{2}$ \;

\Return $\textit{k}$-NN Graph $G$
\caption{GPU-based graph merge (GGM)}
\label{alg:smerge}
\end{algorithm}

Although GGM is simple, it is a critical procedure that allows the \textit{k}-NN graph to be constructed incrementally. As the new data come in, GNND is called to build a sub-graph on the first hand. Thereafter, GGM is called to join this new sub-graph into the existing \textit{k}-NN graph. Similarly, GGM allows the \textit{k}-NN graph to be built on multiple GPUs simultaneously. Given a large-scale dataset, it is partitioned into a number of subsets. The sub-graphs are constructed by GNND for all the subsets on different GPUs. The sub-graphs are later merged one-by-one with GGM. 

In the era of big data, it is possible that the data is too big that the data vectors cannot be held on the memory. In this case, GGM is employed in a different manner to construct the \textit{k}-NN graph.

Firstly,  the large-scale dataset is partitioned into multiple shards. Each shard is sufficiently small that is tractable by one GPU. Thereafter, a \textit{k}-NN graph for each shard is built by GNND and saved back to disk. GGM is called to merge every two sub-graphs of two shards. The merged graph is saved back to disk as two sub-graphs. Each \textit{k}-NN list in either sub-graphs retains the top-\textit{k} neighbors of the corresponding object. Such merge is carried out between sub-graphs pairwisely which ensures any two sub-graphs has been merged once. The resulting sub-graph is a piece of the whole \textit{k}-NN graph, each of which retains the top-\textit{k} neighbors from the whole dataset.

Because only the shards that are being processed need to be kept in the memory, the framework can handle the out-of-GPU-memory datasets. Obviously, multiple merges can be run on multiple GPUs, which makes the framework easy to benefit from multiple GPUs. Although the speed of reading and writing disk is very slow, because the merge operation takes a relatively long time and runs on GPU, we can read and write the disk while merging graphs on GPU. Therefore, with a reasonable arrangement of disk reading and writing, the time spent on large \textit{k}-NN graph construction will be roughly equivalent to the GPU running time.

%% file: exp.tex
\section{Experiments}
\label{sec:experiments}
\begin{table}[t]
  \caption{Summary on Datasets used for Evaluation}
  \begin{tabular}{lcrl} 
  \toprule
  Name & 	$n$ & $d$ & Type \\ 
  \midrule
  SIFT1M~\cite{jegou2010product} & $1{\times}10^6$ & $128$ & SIFT~\cite{lowe2004distinctive} \\ 
  DEEP1M~\cite{babenko2016efficient} & $1{\times}10^6$ & $96$ & Deep \\ 
  GIST1M~\cite{douze2009evaluation} & $1{\times}10^6$ & $960$ & GIST~\cite{douze2009evaluation} \\ 
  GloVe~\cite{pennington2014glove} & $1{\times}10^6$ & $100$ & Text~\cite{pennington2014glove} \\ 
  SIFT10M~\cite{jegou2010product} & $1{\times}10^7$ & $128$ & SIFT \\ 
  SIFT100M~\cite{jegou2010product} & $1{\times}10^8$ & $128$ & SIFT \\ 
  DEEP100M~\cite{babenko2016efficient} & $1{\times}10^8$ & $96$ & Deep \\ 
  DEEP1B~\cite{babenko2016efficient} & $1{\times}10^9$ & $96$ & Deep \\ 
  \bottomrule

  \end{tabular}
  \label{tab:datasets}
\end{table}

\begin{figure}
	\begin{center}
    \includegraphics[width=0.85\linewidth]{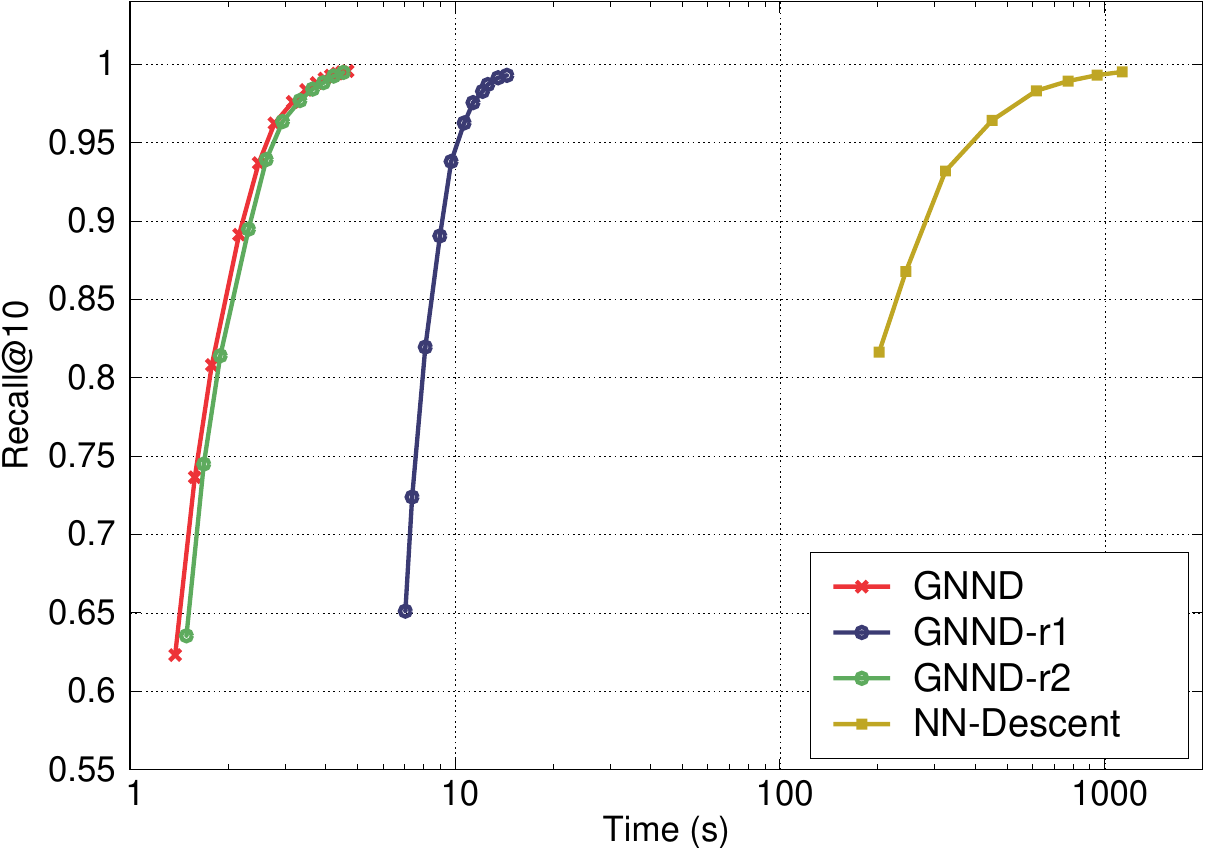}
    \caption{Ablation study over two proposed schemes, namely \textit{selective update} and \textit{multiple spin-locks}. GNND-r1 is a basic GPU implementation without the proposed schemes. GNND-r2 is integrated with \textit{selective update}. GNND is integrated with both schemes. \label{fig:abs}}
\end{center}
\end{figure}

In this section, the performance of GNND is studied in comparison to the representative approaches in the literature. They are FAISS-BF~\cite{johnson2019billion}, GGNN~\cite{groh2019ggnn}, and classic NN-Descent~\cite{weidong}. NN-Descent is treated as the comparison baseline, which is also the only approach that runs on CPU. FAISS-BF constructs \textit{k}-NN graphs on the GPU in a brute-force way. Namely, each sample is compared against the rest of the dataset to get its top-\textit{k} neighbors. GGNN constructs graph index in hierarchy. The bottom layer of the index is a \textit{k}-NN graph for the whole dataset. In addition to FAISS-BF and GGNN, there are other GPU approaches that can be used to construct \textit{k}-NN graph, such as FAISS-IVFPQ~\cite{johnson2019billion} and Sweet KNN~\cite{chen2017sweet}. Unfortunately, they are very slow on million-scale datasets. For instance, the time cost of building indexing alone in FAISS-IVFPQ is already significantly higher than the cost of building a \textit{k}-NN graph by FAISS-BF. The performance of Sweet KNN on such scale dataset is also much poorer than FAISS-BF. Therefore, Sweet KNN is not considered in our comparison. FAISS-IVFPQ is only considered in billion-scale graph construction experiments.

\begin{figure*}[t]
	\begin{center}
	\subfigure[SIFT1M]
    {\includegraphics[width=0.38\linewidth]{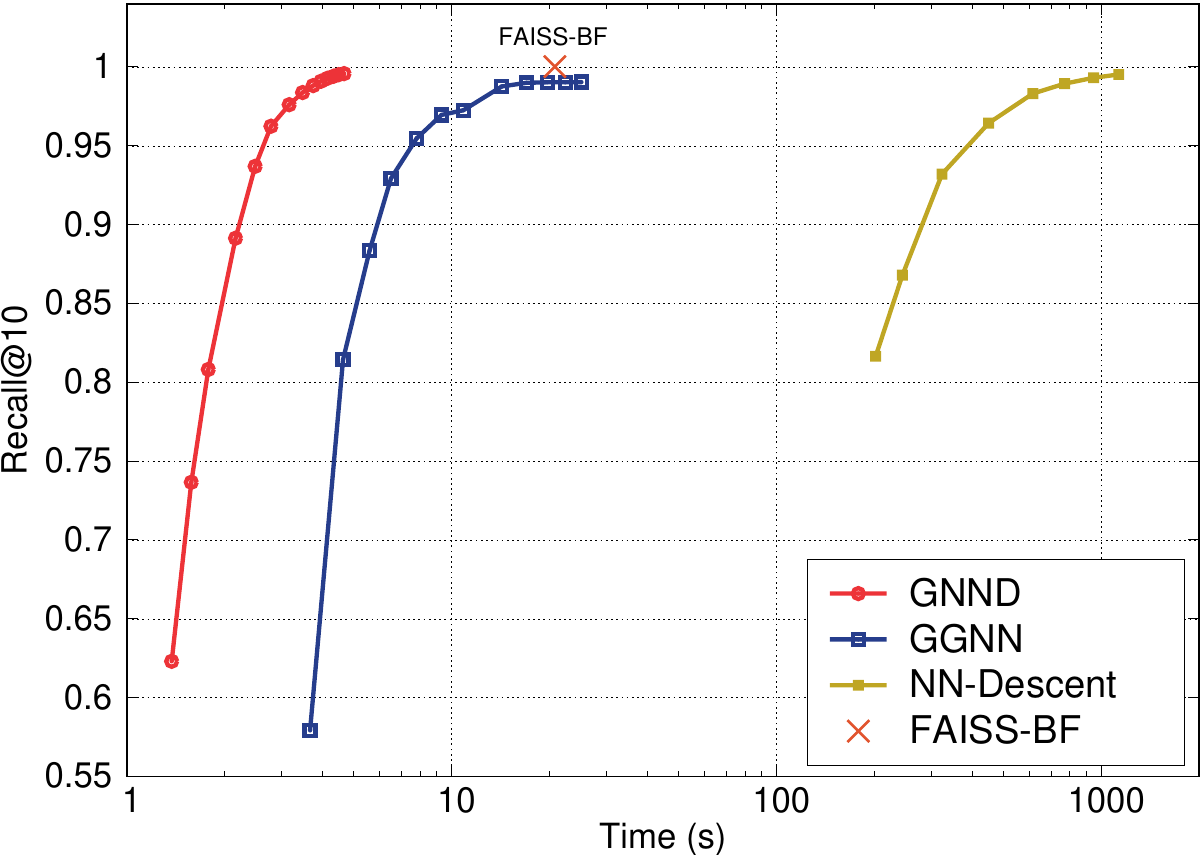}}
    \hspace{0.55in}
	\subfigure[DEEP1M]
    {\includegraphics[width=0.38\linewidth]{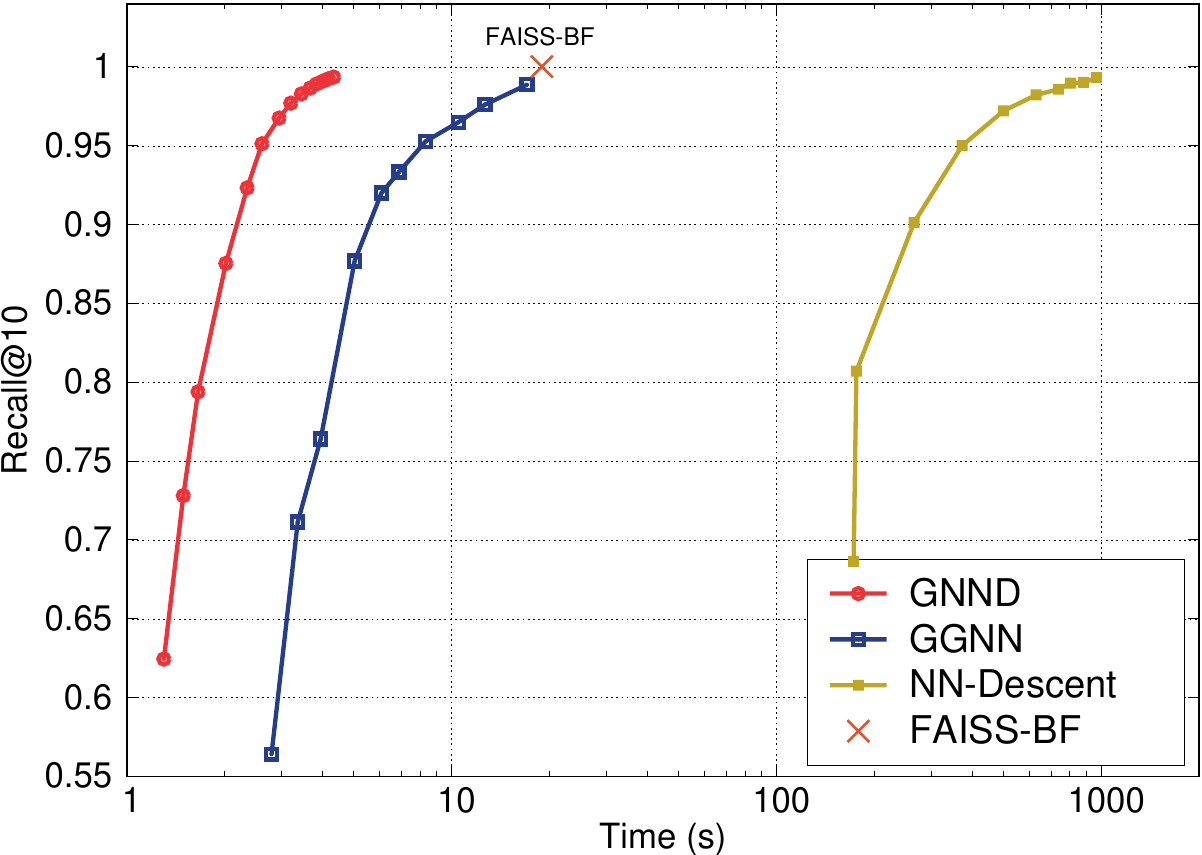}}\\
	\subfigure[GIST1M]
	{\includegraphics[width=0.38\linewidth]{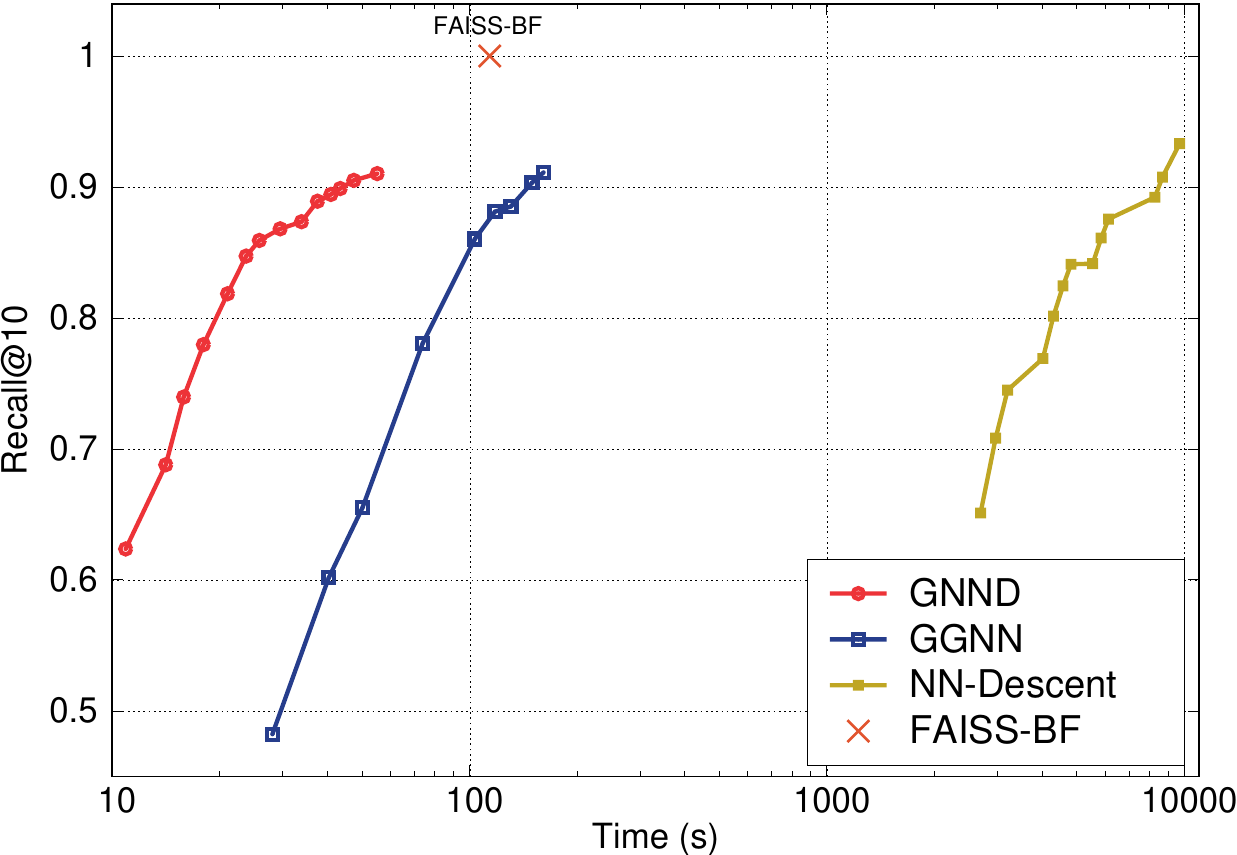}}
    \hspace{0.55in}
	\subfigure[GloVe1M]
	{\includegraphics[width=0.38\linewidth]{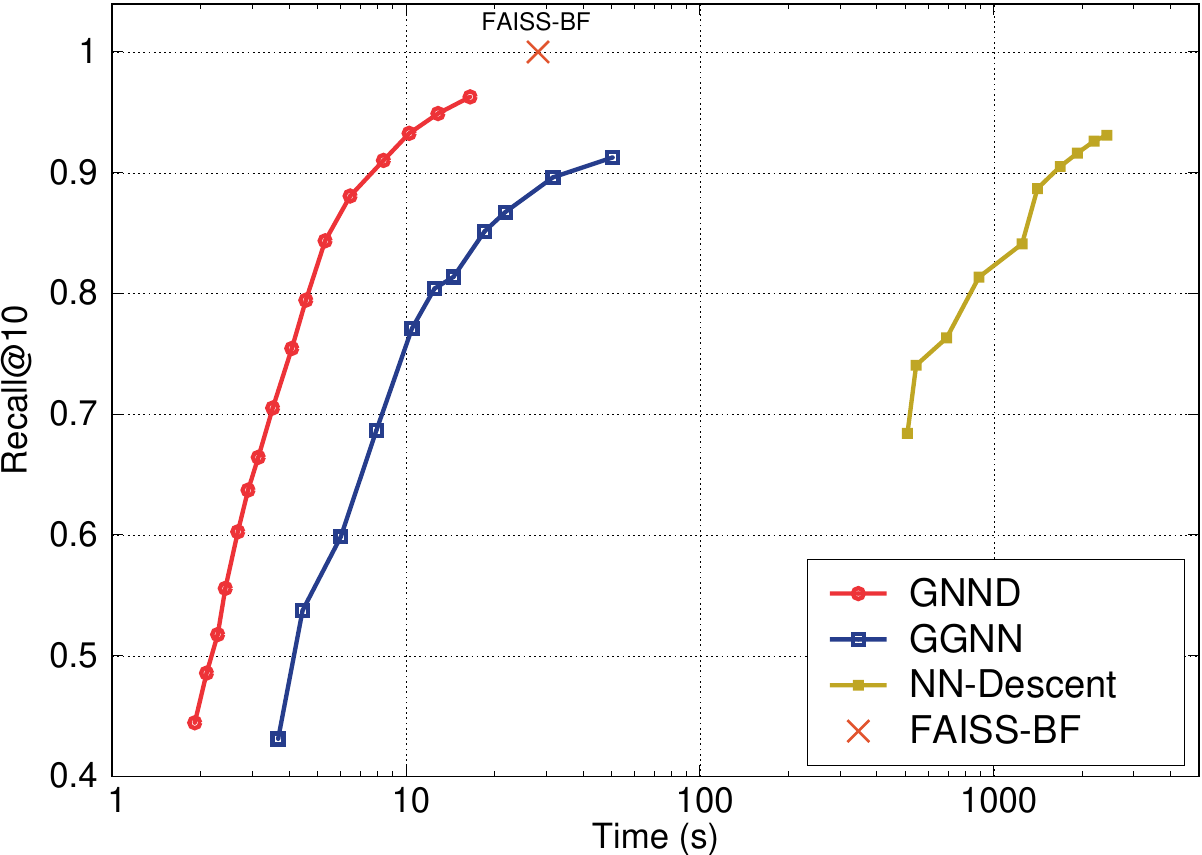}}

	\caption{The \textit{k}-NN graph construction performance on four million-scale datasets. In this experiment, parameters ${k}$ and ${p}$ are varied to produce the curves.}
	\label{fig:exp3_recalls}
\end{center}
\end{figure*}

The experiments are conducted on eight popular ANN benchmark datasets. The scale of these data sets ranges from millions to billions. The brief information of the datasets is summarized in \autoref{tab:datasets}. All the experiments are carried out on a machine with one Intel Xeon Processor W-2123 (3.60Ghz) and 32 GB of memory. The GPU we use is an NVIDIA GeForce RTX 3090.

\subsection{Evaluation Protocol}
The top-\textit{10} recall rate (\textit{Recall@10}) is used to evaluate the quality of the constructed \textit{k}-NN graphs. Given function $R(i,k)$ returns the number of truth-positive neighbors at top-\textit{k} NN list of object $i$, the recall at top-$k$ on the whole graph is given as
\begin{equation}
	Recall@k=\frac{\sum_{i=1}^n{R(i,k)}}{n{\times}k}.
\label{eval:recall}
\end{equation}

\subsection{Ablation Study}

In the first experiment, the speed-up that is achieved by \textit{selectively update} and \textit{multiple spinlock} in GNND is studied. As the strategies in sampling (Section~\ref{sec:smpl}) and the distance calculations on GPU (Section~\ref{sec:crss}) are indispensable parts of our approach, they are integrated with GNND by default. Three runs are carried out in this ablation study. For $\mbox{GNND-r1}$, all the produced neighbor pairs by cross-matching are used to update the corresponding \textit{k}-NN list. Different from insertion performed on CPU, all the calculated distances to a sample $u$ will be sorted by \textit{Batcher’s bitonic sorting network}~\cite{batcher1968sorting}. These sorted distances and their corresponding objects are then merged with the \textit{k}-NN list of $u$. In $\mbox{GNND-r2}$, the proposed \textit{selectively update} is integrated. \autoref{alg:choose_insertion} is called to select the nearest neighbors for insertion. In the third run, namely the full version of GNND, the multiple spinlocks are further set on the \textit{k}-NN lists that allow the insertion to be undertaken in parallel. Please be noted that all of our runs are performed on a single GPU.

In \autoref{fig:abs}, we show the performance from GNND, NN-Descent, $\mbox{GNND-r1}$, and $\mbox{GNND-r2}$. As shown in the figure, more than three times speed-up is observed when \textit{selectively update} is integrated into our approach. Compared to $\mbox{GNND-r2}$, another \textit{5-8\%} speed-up is achieved in GNND as the multiple spin-locks allow the parallel insertion in one \textit{k}-NN list. We also learn that the basic GPU implementation of NN-Descent is more than ten times faster than the classic NN-Descent on CPU. This wide performance gap largely dues to the high parallelization of distance computation on GPU.

\subsection{Million-scale \textit{k}-NN Graph Construction}
In this section, the performance of GNND is compared to NN-Descent~\cite{weidong}, FAISS-BF~\cite{johnson2019billion} and GGNN~\cite{groh2019ggnn} on four million level datasets SIFT1M, DEEP1M, GIST1M, and GloVe1M. For GNND, the parameter \textit{k} is tuned to achieve  a good trade-off between graph quality and efficiency. For GGNN, the number of NN \textit{k} is fixed to \textit{24}, which is in line with the experiments in~\cite{groh2019ggnn} and we find it is efficient on all datasets. The other two parameters slack variable $\tau$ and refinement iterations \textit{t} are tuned to produce \textit{k}-NN graphs of different quality. For DEEP1M and SIFT1M, the parameter \textit{k} is fixed to \textit{64} while parameter \textit{p} is varied from \textit{10} to \textit{32}. For GIST1M and GloVe1M, \textit{k} is varied from \textit{64} to \textit{128} while \textit{p} is varied from \textit{10} to \textit{42}. The results are shown in \autoref{fig:exp3_recalls}.

\autoref{fig:exp3_recalls} shows the experimental results. For SIFT1M, GNND is able to construct a \textit{k}-NN graph with \textit{0.99} recall@\textit{10} in less than \textit{4} seconds, which is \textit{240} times faster than the single-thread NN-Descent. The higher quality of the \textit{k}-NN graph, the wider is the speed gap. FAISS-BF only takes \textit{21} seconds to construct an exact \textit{k}-NN graph for SIFT1M. However, this exhaustive approach is unscalable to datasets with larger scale. For instance, it takes more than \textit{30} minutes to construct an exact \textit{k}-NN graph for SIFT10M. While it only takes around \textit{60} seconds for GNND to reach similar quality. Compared to GGNN, GNND is \textit{2.5} to \textit{5} times faster on the same graph quality level. This observation is consistent across all the datasets. 

\subsection{Billion-Scale \textit{k}-NN Graph Construction}
\begin{figure}[t]
  \centering
  \includegraphics[width=0.85\linewidth]{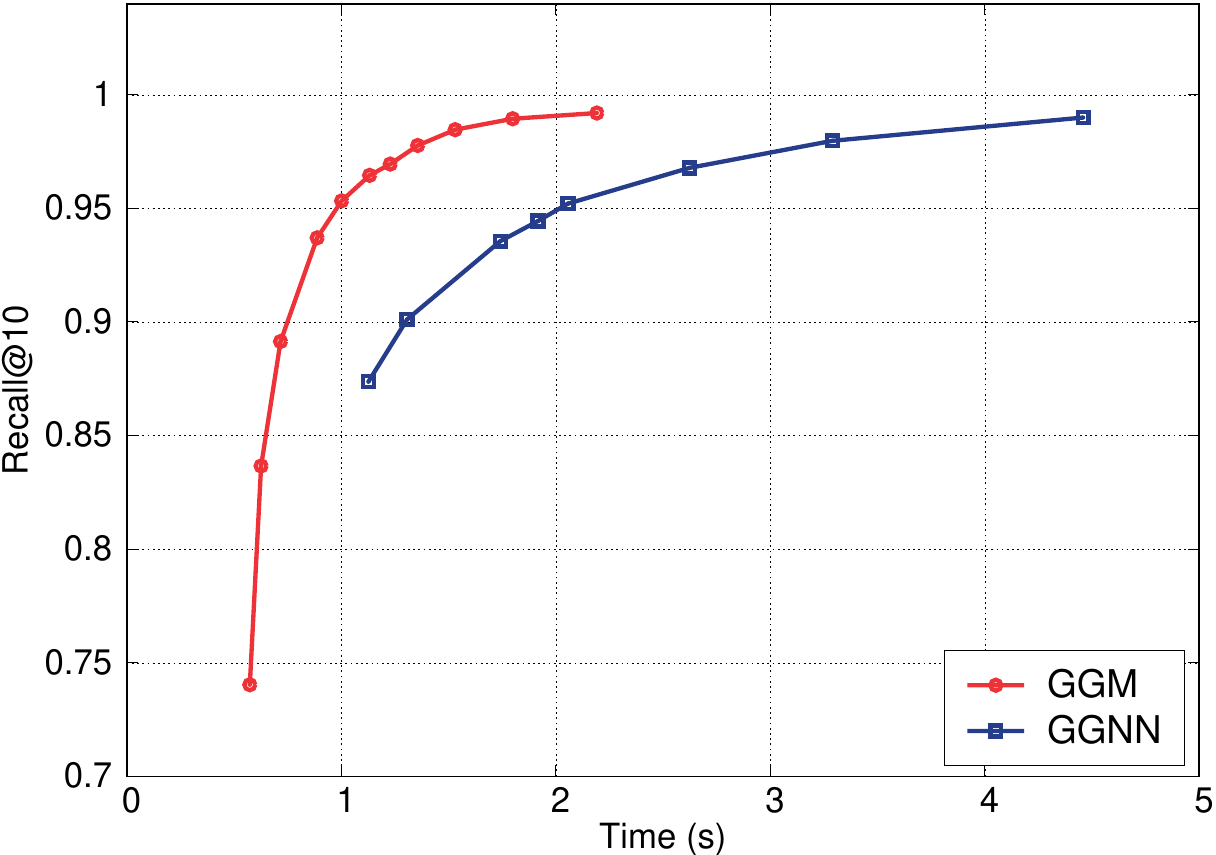}
  \caption{The performance of merging two \textit{k}-NN graphs. The sub-graphs are constructed by GNND. GGM and GGNN are called to merge the two sub-graphs. The time costs of building the sub-graphs by GNND are not counted in the figure.}
  \label{fig:exp2_sift1m_merge}
\end{figure}

In this section, the performance of GNND is further verified on billion level dataset. Before we show the performance on the billion-scale dataset. The performance of GGM is demonstrated, as GGM is the critical procedure in the whole construction pipeline. In our approach, the billion level graph is constructed by pair-wise merging on a group of million level sub-graphs via GGM. In the first experiment, SIFT1M is divided into two 500K datasets. GNND is called to build the \textit{k}-NN graph for each. Thereafter, GGM is called to merge these two sub-graphs. Alternatively, the two sub-graphs are merged by GGNN~\cite{groh2019ggnn} for comparison. The results from GNND and GGNN are shown in \autoref{fig:exp2_sift1m_merge}. It is clear to see GGM is consistently better than GGNN by \textit{5-10\%} in terms of Recall@10. Compared to GGNN, GGM makes the full use of the constructed sub-graphs. GGNN is unable to merge two \textit{k}-NN graphs directly. Instead, \textit{k}-NN search is conducted with samples from one sub-graph against another sub-graph. As a result, only the neighborhood relations of one sub-graph is used during the search.

With the support of GGM, we further show the performance of our approach in building \textit{k}-NN graph for billion-scale datasets. On this scale, it is no longer tractable for the approaches such as GGNN, NN-Descent or FAISS-BF. They either require the whole dataset to be held in memory (e.g., NN-Descent and GGNN) or are too expensive to be undertaken (e.g., FAISS-BF). For this reason,  only FAISS-IVFPQ is considered in our comparison. For our approach, the dataset is partitioned into several hundreds of shards of equal size. Such that we ensure that each shard of the subset is solvable by GNND with a single GPU. After the sub-graphs are constructed by GNND, pairwise merge between the sub-graphs are undertaken as we described in Section~\ref{sec:large-scale}. For FAISS-IVFPQ, a coarse quantizer with $2^{16}$ centroids is trained in advance. Each vector is encoded into 32 bytes by the quantizer. The distance calculation is conducted between these encoded vectors. Due to the low memory consumption, the large-scale dataset could be loaded into memory to construct the \textit{k}-NN graph.

The results from GNND and FAISS-IVFPQ are shown in Table~\ref{tab:bilin}. Since FAISS-IVFPQ cannot construct \textit{k}-NN graph for a billion-scale dataset on a single GPU due to the limited memory capacity of a single GPU, only the results from GNND are presented for SIFT1B and DEEP1B datasets. As shown in the table, the graph quality from FAISS-IVFPQ is considerably lower than our approach. This is mainly because the distance calculations are performed on the compressed data vectors. One could not expect high graph quality even the exhaustive comparison is conducted on the compressed vectors. Another disadvantage of FAISS-IVFPQ is that it is only feasible for $\textit{l}_p$ norms.

\begin{table}
\begin{center}
\caption{Performance of \textit{k}-NN graph Construction on Billion Scale}
\label{tab:bilin}
	\begin{tabular}{|l|lc|lc|} \cline{1-5}
	  & \multicolumn{2}{c|}{GNND} &  \multicolumn{2}{c|}{FAISS-IVFPQ} \\ \hline
	 Dataset  & Time & \textit{Recall@10} & Time & \textit{Recall@10} \\ 
   \cline{1-5}
    SIFT100M & 2,583s & 0.764 & 2,739s & 0.702 \\
    SIFT100M & 3,033s & 0.966 & 4,469s & 0.730 \\
    DEEP100M & 2,364s & 0.767 & 2,331s & 0.705 \\
    DEEP100M & 2,888s & 0.956 & 4,262s & 0.770 \\
    SIFT1B & 77h & 0.955 & - & - \\  
	  DEEP1B & 76h & 0.951 & - & - \\  
   \cline{1-5}
	\end{tabular}
\end{center}

\end{table}

%% file: conc.tex
\section{Conclusion}
\label{sec:conclusion}
In this paper, we have presented an efficient \textit{k}-NN graph construction approach that works on GPUs. It largely follows the major steps in classic NN-Descent which only works well on CPU. In order to adapt to the architecture of GPUs, several optimization schemes are proposed. First of all, the dynamic memory allocation is avoided by fixing the number of samples that are joined in the cross-matching. The distance computations have been specifically designed for different types of sample pairs. A large amount of memory accesses are reduced considerably by \textit{selectively update} without any graph quality degradation. The latency of memory access is further minimized by applying \textit{multiple spinlocks} on a \textit{k}-NN list. All these schemes in combination make full use of the parallelism of the GPU hardware.

In addition, an efficient algorithm is presented to merge two \textit{k}-NN graphs into one. Based on the merge algorithm, it becomes possible to construct \textit{k}-NN graph on multiple GPUs. More importantly, the construction of high-quality \textit{k}-NN graphs has been scaled from million level up-to billion level. The empirical study shows that our approach is 100-250$\times$ faster than single-thread NN-Descent and is 2.5-5$\times$ faster than existing GPU approaches. For out-of-GPU-memory datasets, the quality of the \textit{k}-NN graphs constructed by GNND is significantly higher than the quantization-based approach.